\documentclass[onecolumn]{revtex4}

\usepackage{graphicx}
\usepackage{dcolumn}
\usepackage{bm}

\input epsf
\begin{document}

\title{\Large{Variable $G$ Correction for Dark Energy Model in Higher Dimensional Cosmology}}

\author{\bf Shuvendu Chakraborty$^1$\footnote{shuvendu.chakraborty@gmail.com},
Ujjal Debnath$^2$\footnote{ujjaldebnath@yahoo.com,
ujjal@iucaa.ernet.in} and  Mubasher
Jamil$^3$\footnote{mjamil@camp.nust.edu.pk}}
\affiliation{$^1$Department of Mathematics, Seacom Engineering
College, Howrah, 711 302, India.\\
$^2$Department of Mathematics, Bengal Engineering
and Science University, Shibpur, Howrah-711 103, India.\\
$^3$Center for Advanced Mathematics and Physics (CAMP), National
University of Sciences and Technology (NUST), H-12, Islamabad,
Pakistan.}

\begin{abstract}\vspace{6mm}
 \textbf{Abstract:} In this work, we have considered $N~(=4+d)$-dimensional Einstein
field equations in which 4-dimensional space-time which is
described by a FRW metric and that of the extra $d$-dimensions by
an Euclidean metric. We have calculated the corrections to
statefinder parameters due to variable gravitational constant $G$
in higher dimensional Cosmology. We have considered two special
cases whether dark energy and dark matter interact or not. In a
universe where gravitational constant is dynamic, the variable
$G$-correction to statefinder parameters is inevitable. The
statefinder parameters are also obtained for generalized Chaplygin
gas in the effect of the variation of $G$ correction.\\
\end{abstract}

\maketitle

\section{Introduction}

Cosmological observations obtained by various cosmic explorations of
supernova of type Ia {\cite{c1}}, CMB analysis of WMAP data
{\cite{c2}}, extragalactic explorer SDSS {\cite{c3}} and X-ray
{\cite{c4}} convincingly indicate that the observable universe is
experiencing an accelerated expansion. Although the simplest and
natural solution to explain this cosmic behavior is the
consideration of a cosmological constant \cite{c7}, however it leads
to two relevant problems (namely the ``fine-tuning'' and the
``coincidence'' one). Recently new dynamical nature of dark energy
are considered in the literature, at least in an effective level,
originating from various fields, including a canonical scalar field
(quintessence) \cite{quint}, a phantom field, that is a scalar field
with a negative sign of the kinetic term \cite{phant}, or the
combination of quintessence and phantom in a
unified model named quintom \cite{quintom}.\\

There are some numerous indications that $G$ can be varying and that
there is an upper limit to that variation, with respect to time or
with the expansion of the universe \cite{G4com}. In this connection
the most significant evidences come from the observations of
Hulse-Taylor binary pulsar \cite{Damour,kogan}, helio-seismological
data \cite{guenther}, Type Ia supernova observations \cite{c1}  and
astereoseismological data coming from the pulsating white dwarf star
G117-B15A \cite{Biesiada}; all the above evidences combined lead to
$|\dot{G}/G| \leq 4.10 \times 10^{-11} yr^{-1}$, for $z\lesssim3.5$,
thereby suggesting a mild variation on cosmic level \cite{ray1}. On
a more theoretical level, varying gravitational constant has some
benefits too, for instance it can help alleviating the dark matter
problem \cite{gol}, the cosmic coincidence problem \cite{jamil} and
the discrepancies in Hubble parameter value \cite{ber}. In
literature, a variable gravitational constant has been accommodated
in gravity theories including the Kaluza-Klein \cite{kal},
Brans-Dicke framework
\cite{bd} and scalar-tensor theories \cite{gen}.\\

From the perspective of new gravitational theories including string
theory and braneworld models, (see \cite{string} and references
therein), there are a lot of speculations that there could be extra
dimensions of space (6 extra dimensions of space in the string
theories) besides there are no convincing evidences for their
existence. The size of these dimensions (whether small as Planck
scale or infinitely long like usual dimensions) is still open to
debate. In literature, various theoretical models with extra
dimensions have been constructed to account dark energy \cite{de1}.
Previously some works on variable $G$ correction have been
investigated \cite{Jamil1} to find the statefinder parameters for
several dark energy models. The main motivation of this work is to
investigate the role of statefinder parameters \cite{sahni} (which
can be written in terms of some observable parameters) in higher
dimensional cosmology assuming a varying gravitational constant $G$
for interacting, non-interacting and generalized
Chaplygin gas models.\\

\section{Basic Equations and Solutions}

We consider homogeneous and anisotropic $N$-dimensional space-time
model described by the line element \cite{Paul,Pahwa}

\begin{equation}
ds^{2}=ds^{2}_{FRW}+\sum_{i=1}^{d}b^{2}(t)dx_{i}^{2},
\end{equation}
where $d$ is the number of extra dimensions $(d=N-4)$ and
$ds^{2}_{FRW}$ represents the line element of the FRW metric in
four dimensions is given by

\begin{equation}
ds^{2}_{F R
W}=-dt^{2}+a^{2}(t)\left[\frac{dr^{2}}{1-kr^{2}}+r^{2}(d\theta^{2}+\sin^{2}\theta
d\phi^{2})\right],
\end{equation}
where $a(t)$ and $b(t)$ are the functions of $t$ alone represents
the scale factors of 4-dimensional space time and extra dimensions
respectively. Here  $k ~(=0, ~\pm 1)$  is  the  curvature  index
of the corresponding 3-space, so  that  the  above  model of the
Universe is described  as  flat, closed  and  open respectively.\\

The Einstein's field equations for the above non-vacuum higher
dimensional space-time symmetry are

\begin{equation}
3\left(\frac{\dot{a}^{2}+k}{a^{2}}\right)=\frac{\ddot{D}}{D}-\frac{d^{2}}{8}
\frac{\dot{b}^{2}}{b^{2}}+\frac{d}{8} \frac{\dot{b}^{2}}{b^{2}}+8
\pi G \rho
\end{equation}

\begin{equation}
2\frac{\ddot{a}}{a}+\frac{\dot{a}^{2}+k}{a^{2}}=\frac{\dot{a}}{a}\frac{\dot{D}}{D}+\frac{d^{2}}{8}
\frac{\dot{b}^{2}}{b^{2}}-\frac{d}{8} \frac{\dot{b}^{2}}{b^{2}}-8
\pi G p
\end{equation}
and
\begin{equation}
\frac{\ddot{b}}{b}+3\frac{\dot{a}}{a}\frac{\dot{b}}{b}=-\frac{\dot{D}}{D}\frac{\dot{b}}{b}+\frac{\dot{b}^{2}}{b^{2}}-\frac{8
\pi G p}{2}
\end{equation}

where $\rho$ and $p$ are energy density and isotropic pressure of
the fluid filled in the universe respectively. We choose,
$D^{2}=b^{d}(t)$, so we have
$\frac{\dot{D}}{D}=\frac{d}{2}\frac{\dot{b}}{b}$ and
$\frac{\ddot{D}}{D}=\frac{d}{2}\frac{\ddot{b}}{b}+\frac{d^{2}-2d}{4}\frac{\dot{b}^{2}}{b^{2}}$.
Hence the equations (1), (2) and (3) become

\begin{equation}
3\left(\frac{\dot{a}^{2}+k}{a^{2}}\right)=\frac{d}{2}\frac{\ddot{b}}{b}+\frac{d^{2}-2
d }{4}\frac{\dot{b}^{2}}{b^{2}}-\frac{d^{2}}{8}
\frac{\dot{b}^{2}}{b^{2}}+\frac{d}{8} \frac{\dot{b}^{2}}{b^{2}}+8
\pi G \rho,
\end{equation}

\begin{equation}
2\frac{\ddot{a}}{a}+\frac{\dot{a}^{2}+k}{a^{2}}=\frac{d}{2}\frac{\dot{a}}{a}\frac{\dot{b}}{b}+\frac{d^{2}}{8}
\frac{\dot{b}^{2}}{b^{2}}-\frac{d}{8} \frac{\dot{b}^{2}}{b^{2}}-8
\pi G p,
\end{equation}
and
\begin{equation}
\frac{\ddot{b}}{b}+3\frac{\dot{a}}{a}\frac{\dot{b}}{b}=-\frac{\dot{d}}{2}\frac{\dot{b^{2}}}{b^{2}}+\frac{\dot{b}^{2}}{b^{2}}-\frac{8
\pi G p}{2}.
\end{equation}

Defining $H_{1}=\frac{\dot{a}}{a}$, $H_{2}=\frac{\dot{b}}{b}$ we
have from (6) to (8),

\begin{equation}
3 H_{1}^{2}+3 \frac{k}{a^{2}}-\frac{d}{2}\dot{H}_{2}+\left(
-\frac{d}{8}-\frac{d^{2}}{8}\right)H_{2}^{2}=8 \pi G \rho,
\end{equation}

\begin{equation}
3 H_{1}^{2}+ \frac{k}{a^{2}}+2
\dot{H}_{1}-\frac{d}{2}H_{1}H_{2}+\left(
\frac{d}{8}-\frac{d^{2}}{8}\right)H_{2}^{2}=-8 \pi G p,
\end{equation}

\begin{equation}
\dot{H}_{2}+3 H_{1} H_{2}-\frac{d}{2}H_{2}^{2}=-\frac{8 \pi Gp}{2}.
\end{equation}
Now consider the universe is filled with the dark matter (with
negligible pressure) and dark energy. Assuming $p=\omega \rho_{x}$,
$\rho= \rho_{m}+\rho_{x}$, where $\rho_{m}$ and $\rho_{x}$ are the
energy densities of dark matter and dark energy respectively,
$\omega$ is the equation of state parameter for dark energy. Note
that $\omega$ is a dynamical time dependent parameter and will be
useful in later calculations. Now eliminating $\dot{H}_{1}$,
$\dot{H}_{2}$ from (9), (10) and (11) we have,

\begin{equation}
24 k= a^{2}\left(-24 H_{1}^{2}-12 d H_{1}H_{2}-(d-1)d H_{2}^{2}+16
\pi G (4 \rho_{m}+(4-d \omega)\rho_{x})\right)
\end{equation}

This equation can be written as

\begin{equation}
(d+3)^{2}H^{2}+3H_{1}^{2}-\frac{d(d-1)}{2}H_{2}^{2}+\frac{12
k}{a^{2}}=32 \pi G \rho_{m}+(4-d \omega)8 \pi G \rho_{x}
\end{equation}

where $H$ is the Hubble parameter defined by
$H=\frac{1}{d+3}\left(3 H_{1}+d H_{2}\right)$. This can be written
as
\begin{equation}
\Omega+\frac{3}{(d+3)^{2}}\Omega_{1}-\frac{d(d-1)}{2}\Omega_{2}+\frac{12
}{(d+3)^{2}}\Omega_{k}=\frac{4}{d+3}\Omega_{m}+\frac{4-d
\omega}{d+3}\Omega_{x}
\end{equation}

where $\Omega_{1}=\frac{H_{1}^{2}}{H^{2}}$,
$\Omega_{2}=\frac{H_{2}^{2}}{H^{2}}$ are dimensionless parameters
and $\Omega_{m}=\frac{8 \pi G \rho_{m}}{(d+3)H^{2}}$,
$\Omega_{x}=\frac{8 \pi G \rho_{x}}{(d+3)H^{2}}$ are fractional
density parameters, $\Omega_{k}=\frac{k}{a^{2}H^{2}}$ is another
dimensionless parameter, represents the contribution in the energy
density from the spatial curvature and $\Omega$ is the total
density parameter. Now solving (9), (10) and (11) we have the
solutions of $\dot{H}_{1}$ and $\dot{H}_{2}$ as

\begin{equation}
\dot{H}_{1}=\frac{1}{24}\left(-48
H_{1}^{2}+d(d-1)H_{2}^{2}-\frac{24 k}{a^{2}}+8 \pi G(4
\rho_{m}-(-4+(d+12)\omega)\rho_{x})\right)
\end{equation}
and
\begin{equation}
\dot{H}_{2}=\frac{2}{d}\left( 3
H_{1}^{2}-\frac{1}{8}d(d+1)H_{2}^{2}+\frac{3 k}{a^{2}}-8 \pi
G(\rho_{m}+\rho_{x}) \right)
\end{equation}

The deceleration parameter $q=-1-\frac{\dot{H}}{H^{2}}$ is given
by in terms of dimensionless parameters
\begin{equation}
q=-1-\frac{3}{d+3}\Omega_{k}+\frac{d}{8}\Omega_{2}+\frac{3}{2}\Omega_{m}+\frac{12+12
\omega+d\omega }{8}\Omega_{x}
\end{equation}

and the derivative of deceleration parameter is obtained as

\begin{equation}
\dot{q}=\frac{3}{d+3}H \Omega_{k}\left(2
\sqrt{\Omega_{1}}-2(q+1)\right)+\frac{d}{8}\dot{\Omega}_{2}+\frac{3}{2}\dot{\Omega}_{m}+\frac{12+12
\omega+d \omega
}{8}\dot{\Omega}_{x}+\frac{d+12}{8}\dot{\omega}\Omega_{x}
\end{equation}

Also from (14) we obtain the expression of the total density
parameter in the form

\begin{equation}
\Omega=\frac{4}{d+3}\frac{\rho}{\rho_{cr}}-\frac{d
\omega}{d+3}\Omega_{x}-\frac{12
}{(d+3)^{2}}\Omega_{k}-\frac{3}{(d+3)^{2}}\Omega_{1}+\frac{d(d-1)}{2}\Omega_{2}
\end{equation}
Now define the critical density,
\begin{equation}
\rho_{cr}=\frac{3 H^{2}}{8 \pi G(t)}~~~~~~ \text{which gives after
differentiation}~~ \dot{\rho}_{cr}=\rho_{cr}\left(
2\frac{\dot{H}}{H}-\frac{\dot{G}}{G}\right)
\end{equation}
which implies
\begin{equation}
\dot{\rho}_{cr}=-H \rho_{cr}(2(1+q)+\triangle G)
\end{equation}

where, $\triangle G\equiv \frac{G'}{G}, \dot{G}=H G'$ (prime denotes
differentiation with respect to $x\equiv\ln a$). The benefit of the
previous rule $\dot{G}=H G'$ relates the variations in $G$ with
respect to time $\dot G$ and the expansion of the universe $G'$.
Differentiating (19) we have

\begin{equation}
\dot{\Omega}=\frac{4}{d+3}\frac{\dot{\rho}}{\rho_{cr}}+\frac{4
H(2(1+q)+\triangle G)}{d+3}\frac{\rho}{\rho_{cr}}-\frac{24
H}{(d+3)^{2}}\Omega_{k}(\sqrt{\Omega_{1}}-(q+1))-\frac{3}{(d+3)^{2}}\dot{\Omega_{1}}+\frac{d(d-1)}{2}\dot{\Omega}_{2}-\frac{d}{d+3}\omega
\dot{\Omega}_{x}-\frac{d}{d+3}\dot{\omega}\Omega_{x}
\end{equation}
where $\dot{\Omega_{1}}$ and $\dot{\Omega}_{2}$ are given by

\begin{eqnarray*}
\dot{\Omega_{1}}=H \left[ \Omega_{1}^{1/2}\Omega_{k}-3
\Omega_{1}^{2}-\frac{3d}{2(d+3)}\Omega_{1}^{3/2}\Omega_{2}^{1/2}+\frac{d(d-1)}{8}\Omega_{1}^{1/2}\Omega_{2}-(d+3)\omega
\Omega_{1}^{1/2}\Omega_{x}-\frac{9}{d+3}\Omega_{1}
\Omega_{k}+\frac{d}{2}\Omega_{1}\Omega_{2}^{1/2}\right.
\end{eqnarray*}
\begin{equation}
\left.+\frac{d(d+7)}{8(d+3)}\Omega_{1} \Omega_{2}+4\Omega_{1}
\Omega_{m}+(4+3\omega)\Omega_{1} \Omega_{x} \right]
\end{equation}

\begin{eqnarray*}
\dot{\Omega}_{2}=H\left[
\frac{12}{d}\Omega_{1}^{3/2}-\frac{3}{d+3}\Omega_{1}^{3/2}\Omega_{2}^{1/2}-\frac{d+1}{2}
\Omega_{1}^{1/2}\Omega_{2}-\frac{3d}{2(d+3)}\Omega_{1}^{3/2}\Omega_{2}+\frac{d(d+7)}{8(d+3)}\Omega_{1}^{1/2}\Omega_{2}^{3/2}+\frac{12}{d}
\Omega_{1}^{1/2}\Omega_{k}\right.
\end{eqnarray*}
\begin{equation}
\left.-\frac{9}{d+3}\Omega_{1}^{1/2}\Omega_{2}^{1/2}\Omega_{k}-\frac{4(d+3)}
{d}\Omega_{1}^{1/2}(\Omega_{m}+\Omega_{x})+\Omega_{1}^{1/2}\Omega_{2}^{1/2}\left(4\Omega_{m}+(4+3\omega)\Omega_{x}\right)\right]
\end{equation}

The trajectories in the \{$r,s$\} plane corresponding to different
cosmological models depict qualitatively different behaviour. The
statefinder diagnostic along with future SNAP observations may
perhaps be used to discriminate between different dark energy
models. The above statefinder diagnostic pair for cosmology are
constructed from the scale factor $a$. The statefinder parameters
are given by \cite{sahni}
$$
r=\frac{\dddot{a}}{aH^{2}}~,~~s=\frac{r-1}{3(q-1/2)}
$$

Now we obtain the expressions for $r$ and $s$ as follows

\begin{eqnarray*}
r=\frac{d^{2}}{32}\Omega_{2}^{2}-\frac{d}{8}\frac{\dot{\Omega}_{2}}{H}+(12+(d+12)\omega)\Omega_{x}\left(-\frac{3}{8}+\frac{3}{4}\Omega_{m}
-\frac{3}{4(d+3)}\Omega_{k}+\frac{d}{16}\Omega_{2}-\frac{1}{8}\frac{\dot{\Omega}_{x}}{H}
\right)-\frac{3}{d+3}\Omega_{k}(3\Omega_{m}-3+2\Omega_{1}^{1/2})
\end{eqnarray*}
\begin{equation}
-\frac{3d}{4(d+3)}\Omega_{2}
\Omega_{k}+\frac{3d}{8}\Omega_{2}(2\Omega_{m}-1)+4\Omega_{m}^{2}-\frac{3}{2}\frac{\dot{\Omega}_{m}}
{H}-\frac{9}{2}\Omega_{m}+\frac{1}{32}[(12+(d+12)\omega)^{2}\Omega_{x}^{2}]-\frac{(d+12)}{8}\frac{\dot{\omega}}{H}\Omega_{x}+1
\end{equation}

\begin{eqnarray*}
s=\frac{8(d+3)}{3[d(d+3)\Omega_{2}-24
\Omega_{k}+(d+3)\left(-12+12\Omega_{m}+(12+(d+12)\omega)\Omega_{x}\right)]
}\left[\frac{d^{2}}{32}\Omega_{2}^{2}-\frac{d}{8}\frac{\dot{\Omega}_{2}}{H}~~~~~~~~~~~~~~~~~~~~~~~~~~~~~~~~~~~~~\right.
\end{eqnarray*}
\begin{eqnarray*}
+(12+(d+12)\omega)\Omega_{x}\left(-\frac{3}{8}+\frac{3}{4}\Omega_{m}
-\frac{3}{4(d+3)}\Omega_{k}+\frac{d}{16}\Omega_{2}-\frac{1}{8}\frac{\dot{\Omega}_{x}}{H}
\right)-\frac{3}{d+3}\Omega_{k}(3\Omega_{m}-3+2\Omega_{1}^{1/2})~~~~~~~~~~~~~~~~~~~~
\end{eqnarray*}
\begin{equation}
\left.-\frac{3d}{4(d+3)}\Omega_{2}
\Omega_{k}+\frac{3d}{8}\Omega_{2}(2\Omega_{m}-1)+4\Omega_{m}^{2}-\frac{3}{2}\frac{\dot{\Omega}_{m}}
{H}-\frac{9}{2}m+\frac{1}{32}\left((12+(d+12)\omega)^{2}\Omega_{x}^{2}\right)-\frac{(d+12)}{8}\frac{\dot{\omega}}{H}\Omega_{x}\right]
\end{equation}

This is the expressions for $\{r,s\}$ parameters in terms of
fractional densities of dark energy model in higher dimensional
cosmology for closed (or open) universe where the derivative of
the density parameters i.e., $\dot{\Omega}_{1}$ and
$\dot{\Omega}_{2}$ are given in equation (23) and (24). Now in the
following subsections, we shall analyze the statefinder parameters
for the non-interacting and interacting dark energy models.\\

\subsection{Non-interacting Dark Energy Model}

In this subsection we study the model of non-interacting case
where the dark energy and dark matter do not interact with each
other. We assume that dark matter and dark energy are separately
conserved. So the continuity equation for cold dark matter is
$\dot{\rho}_{m}+(d+3) H \rho_{m}=0$ and for dark energy is
$\dot{\rho}_{x}+(d+3) H (1+\omega)\rho_{x}=0$. So solving (22) for
two different cases we have the expressions of $\dot{\Omega}_{m}$
and $\dot{\Omega}_{x}$ as:

\begin{eqnarray*}
\dot{\Omega}_{m}=\frac{1}{2(d+3)(d+3+d
\omega)}\left[-6\dot{\Omega_{1}}+d(d-1)(d+3)^{2}\dot{\Omega}_{2}+2\left(24H(q+1-\Omega_{1}^{1/2})\Omega_{k}+4H((d+3)(2q-1-d+\triangle
G)\right.\right.
\end{eqnarray*}
\begin{equation}
\left.\left.+d(2q+\triangle G+2
)\omega+d(d+3)\omega^{2})\Omega_{m}-d(4H(2q+2+\triangle
G)\omega+(d+3)\dot{\omega})\Omega_{x}\right)\right]
\end{equation}

\begin{eqnarray*}
\dot{\Omega}_{x}=\frac{1}{2(d+3)(d+3+d
\omega)}\left[-6\dot{\Omega_{1}}+d(d-1)(d+3)^{2}\dot{\Omega}_{2}+2
\left(24H(q+1-\Omega_{1}^{1/2})\Omega_{k}-(d+3)(4(d+3)H(\omega+1)\Omega_{m}\right.\right.
\end{eqnarray*}
\begin{equation}
\left. \left.+(-4H(2q+2+\triangle G )+d
\dot{\omega})\Omega_{x})\right)\right]
\end{equation}

In the equations (25) and (26), we have calculated the general
expressions of the statefinder parameters $\{r, s\}$. In this
non-interacting dark energy model, the above parameters are also
same where the $\dot{\Omega}_{m}$ and $\dot{\Omega}_{x}$ are given
by the equations (27) and (28).\\

\subsection{Interacting Dark energy Model}

In this subsection we study the model of interacting case where
the dark energy and dark matter are interact with each other.
These models describe an energy flow between the components i.e.
they not separately conserved. According to recent observational
data of Supernovae and CMB the present evolution of the Universe
permit the energy transfer decay rate proportional to present
value of the Hubble parameter. Many authors have widely studied
this interacting model. In Pavon and Zimdahl \cite{Pavon} state
that the unknown nature of dark energy and dark matter make no
contradiction about their mutual interaction. In Zhang and Olivers
et al \cite{Zhang} showed that the theoretical interacting model
are consistent with the type
Ia supernova and CMB observational data.\\

Here we assume that the dark energy and dark matter are
interacting with each other, so the continuity equations of dark
matter and dark energy become

\begin{equation}
\dot{\rho}_{m}+(d+3) H \rho_{m}=Q
\end{equation}
and
\begin{equation}
\dot{\rho}_{x}+(d+3) H (1+\omega)\rho_{x}=-Q
\end{equation}
where $Q$ is is the interacting term which is a arbitrary
function. This interacting term determine the direction of the
energy flow both sides of the dark matter and dark energy. In
general this term can be choose as a function of different
cosmological parameters like Hubble parameter and dark energy or
dark matter density. in this work we choose $Q=(d+3) \delta H
\rho_{x}$ where $\delta $ is a couple constant. The positive
$\delta$ represents the energy transfer from dark energy to dark
matter. If $\delta=0$ the above model transfer to non-interacting
case. Here negative $\delta$ is not considered as it can violate
the thermodynamical laws of the universe. So the
$\dot{\Omega}_{m}$ and $\dot{\Omega}_{x}$ are given by the
equations
\begin{eqnarray*}
\dot{\Omega}_{m}=\frac{1}{2(d+3)(d+3+d
\omega)}\left[-6\dot{\Omega_{1}}+d(d-1)(d+3)^{2}\dot{\Omega}_{2}+2\left(24H(
q+1-\Omega_{1}^{1/2})\Omega_{k}+4H((d+3)( 2q-1-d+\triangle
G)\right.\right.
\end{eqnarray*}
\begin{equation}
\left.\left.+d(2+\triangle
G+2q)\omega+d(d+3)\omega^{2})\Omega_{m}+(4H(-d(2q+2+\triangle
G)\omega+(d+3)\delta(d+3+2 d
\omega))-d(d+3)\dot{\omega})\Omega_{x}\right)\right]
\end{equation}

\begin{eqnarray*}
\dot{\Omega}_{x}=\frac{1}{2(d+3)( d+3+d
\omega)}\left[-6\dot{\Omega_{1}}+d(d-1)(d+3)^{2}\dot{\Omega}_{2}+2
\left(24H(q+1-\Omega_{1}^{1/2})\Omega_{k}\right.\right.~~~~~~~~~~~~~~~~~~~~~~~~~~
\end{eqnarray*}
\begin{equation}
\left.
\left.-(d+3)(4H(d+3)(1+\omega)\Omega_{m}+(-4H(2q+2+\triangle G
)+4(d+3) H \delta+d \dot{\omega})\Omega_{x})\right)\right]
\end{equation}

In the equations (25) and (26), we have calculated the general
expressions of the statefinder parameters $\{r, s\}$. In this
interacting dark energy model, the above parameters are also same
where the $\dot{\Omega}_{m}$ and $\dot{\Omega}_{x}$ are given
by the equations (31) and (32).\\

\section{Generalized Chaplygin gas}
It is well known to everyone that Chaplygin gas provides a
different way of evolution of the universe and having behaviour at
early time as presureless dust and as cosmological constant at
very late times, an advantage of  generalized Chaplygin gas (GCG),
that is it unifies dark energy and dark matter into a single
equation of state. This model can be obtained from generalized
version of the Born-Infeld action. The equation of state for
generalized Chaplygin gas is \cite{Gorini}
\begin{equation}
p_{x}=-\frac{A}{\rho_{x}^{\alpha}}
\end{equation}
 where $0<\alpha<1$ and $A>0$ are constants. Inserting the above equation of state (33) of the GCG into the non-interacting energy conservation equation
 we have
\begin{equation}
\rho_{x}=\left[A+\frac{B}{(a^{3}b^{d})^{(\alpha+1)}}\right]^{\frac{1}{\alpha+1}}
\end{equation}
where $B$ is an integrating constant.
\begin{equation}
\omega=-A \left(A +
\frac{B}{(a^{3}b^{d})^{(\alpha+1)}}\right)^{-1}
\end{equation}
 Differentiating (35) we have

\begin{equation}
 \frac{\dot{\omega}}{H}=-(d+3) A B (1 + \alpha) \frac{1}{(a^{3}b^{d})^{(\alpha+1)}}
  \left(A + \frac{B}{(a^{3}b^{d})^{(\alpha+1)}}\right)^{-2}
\end{equation}
Now putting (36) in (25) and (26), we have

\begin{eqnarray*}
r=\frac{d^{2}}{32}\Omega_{2}^{2}-\frac{d}{8}\frac{\dot{\Omega}_{2}}{H}-\frac{9}{2}\Omega_{m}+\frac{9}{2}\Omega_{m}^{2}-\frac{3}{2}\frac{\dot{\Omega}_{m}}{H}
+\frac{9}{d+3}\Omega_{k}-\frac{6}{d+3}\Omega_{1}^{1/2}\Omega_{k}-\frac{3d}{d+3}\Omega_{2}
\Omega_{k}
-\frac{3d}{8}\Omega_{2}~~~~~~~~~~~~~~~~~~~~~~~~~~~~~~~~~~~~~~~
\end{eqnarray*}
\begin{eqnarray*}
-\frac{36B^{2}-3a^{6}A^{2}b^{2d}(a^{3}b^{d})^ {2\alpha}d-A
B(a^{3}b^{d})^{\alpha+1}(d(d+18)+(d+3)(d+12)\alpha)}{8(A(a^{3}b^{d})^{\alpha+1}+B)^{2}}\Omega_{x}+\frac{1}{32}\left(-12+\frac{A(d+12)}{A+
(a^{3}b^{d})^{-\alpha-1}B}\right)^{2}\Omega_{x}^{2}
\end{eqnarray*}
\begin{equation}
+\left(12-\frac{A(d+12)}{A+
(a^{3}b^{d})^{-\alpha-1}B}\right)\left(\frac{3}{4}\Omega_{m}
\Omega_{x}-\frac{\dot{\Omega}_{x}}{8H}\right)+\left(
\frac{B(d+12)\Omega_{x}}{A(a^{3}b^{d})^{\alpha+1}+B}+12
\Omega_{m}-d
\Omega_{x}\right)\left(\frac{d}{16}\Omega_{2}-\frac{3}{4(d+3)}\Omega_{k}\right)+1
 \end{equation}

 \begin{eqnarray*}
s=\frac{1}{-\frac{9}{2}+\frac{3d\Omega_{2}}{8}-\frac{9\Omega_{k}}{d+3}+\frac{9\Omega_{m}}{2}-\frac{3d\Omega_{x}}{8}+\frac{3}{8}
\frac{B(d+12)\Omega_{x}}{A
(a^{3}b^{d})^{\alpha+1}+B}}\left[\frac{d^{2}}{32}\Omega_{2}^{2}-\frac{d}{8}
\frac{\dot{\Omega}_{2}}{H}-\frac{9}{2}\Omega_{m}+4\Omega_{m}^{2}-\frac{3}{2}\frac{\dot{\Omega}_{m}}{H}
+\frac{9}{d+3}\Omega_{k}-\frac{6}{d+3}\Omega_{1}^{1/2}
\Omega_{k}-\frac{3d}{8}\Omega_{2} \right.
\end{eqnarray*}
\begin{eqnarray*}
-\frac{36B^{2}-3a^{6}A^{2}b^{2d}(a^{3}b^{d})^ {2\alpha}d-A
B(a^{3}b^{d})^{\alpha+1}(d(d+18)+(d+3)(d+12)\alpha)}{8(A(a^{3}b^{d})^{\alpha+1}+B)^{2}}\Omega_{x}+\frac{1}{32}\left(-12+\frac{A(d+12)}{A+
(a^{3}b^{d})^{-\alpha-1}B}\right)^{2}\Omega_{x}^{2}
\end{eqnarray*}
\begin{equation}
\left.-\frac{3d}{d+3}\Omega_{2}+\left(12-\frac{A(d+12)}{A+
(a^{3}b^{d})^{-\alpha-1}B}\right)\left(\frac{3}{3}\Omega_{m}
\Omega_{x}-\frac{\dot{\Omega}_{x}}{8H}\right)+\left(
\frac{B(d+12)\Omega_{x}}{A(a^{3}b^{d})^{\alpha+1}+B}+12
\Omega_{m}-d
\Omega_{x}\right)\left(\frac{d}{16}\Omega_{2}-\frac{3}{4(d+3)}\Omega_{k}\right)\right]
 \end{equation}
These are the expressions for $\{r,s\}$ parameters in terms of
fractional densities for non-interacting case of generalized
Chaplygin gas model in higher dimensional Cosmology, where
$\dot{\Omega}_{m}$ and $\dot{\Omega}_{x}$ are given
by the equations (27) and (28).\\

Again inserting the equation of state (33) of the GCG into the
interacting energy conservation equation (30) we have

\begin{equation}
\rho_{x}=\left[\frac{A}{\delta+1}+\frac{B}{(\delta+1)(a^{3}b^{d})^{(\delta+1)(\alpha+1)}}\right]^{\frac{1}{\alpha+1}}
\end{equation}
and
\begin{equation}
\omega=-A \left(\frac{A}{\delta+1} +
\frac{B}{(\delta+1)(a^{3}b^{d})^{(\delta+1)(\alpha+1)}}\right)^{-1}
\end{equation}

Differentiating (40) we have

\begin{equation}
 \frac{\dot{\omega}}{H}=-(d+3) A B (1 + \alpha) \frac{1}{(a^{3}b^{d})^{(\delta+1)(\alpha+1)}}
 \left(\frac{A}{\delta+1} +
\frac{B}{(\delta+1)(a^{3}b^{d})^{(\delta+1)(\alpha+1)}}\right)^{-2}
\end{equation}
Now putting (41) in (25) and (26), we have

\begin{eqnarray*}
r=\frac{d^{2}}{32}\Omega_{2}^{2}-\frac{d}{8}\frac{\dot{\Omega}_{2}}{H}-\frac{9}{2}\Omega_{m}+\frac{9}{2}\Omega_{m}^{2}-\frac{3}{2}\frac{\dot{\Omega}_{m}}{H}
+\frac{9}{d+3}\Omega_{k}-\frac{6}{d+3}\Omega_{1}^{1/2}\Omega_{k}-\frac{3d}{d+3}\Omega_{2}
\Omega_{k}
-\frac{3d}{8}\Omega_{2}-\frac{9}{d+3}\Omega_{k}\Omega_{m}+\frac{3d}{4}\Omega_{2}\Omega_{m}
\end{eqnarray*}
\begin{eqnarray*}
+\frac{A
B(a^{3}b^{d})^{(1+\alpha)(1+\delta)}(d+3)(d+12)(1+\alpha)(1+\delta)^{2}}{8(A(a^{3}b^{d})^{(1+\alpha)(1+\delta)}+B)^{2}}
+\left(\frac{d}{16}\Omega_{2}+\frac{3}{4}\Omega_{m}-\frac{3}{4(d+3)}\Omega_{k}-\frac{3}{8}\right)
\end{eqnarray*}
\begin{equation}
\times \left(
-12+\frac{A(d+12)(1+\delta)}{A+B(a^{3}b^{d})^{-(1+\alpha)(1+\delta)}}
\right)-\left(
-12+\frac{A(d+12)(1+\delta)}{A+B(a^{3}b^{d})^{-(1+\alpha)(1+\delta)}}
\right)\frac{\dot{\Omega}_{x}}{8H}+1
 \end{equation}

 \begin{eqnarray*}
s=\frac{1}{3d\Omega_{2}-\frac{72}{d+3}\Omega_{k}+36\Omega_{m}-36+3\left(
-12+\frac{A(d+12)(1+\delta)}{A+B(a^{3}b^{d})^{-(1+\alpha)(1+\delta)}}
\right)\Omega_{x}
}\left[\frac{d^{2}}{32}\Omega_{2}^{2}-\frac{d}{8}\frac{\dot{\Omega}_{2}}{H}-\frac{9}{2}\Omega_{m}+\frac{9}{2}\Omega_{m}^{2}-\frac{3}{2}
\frac{\dot{\Omega}_{m}}{H}+\frac{9}{d+3}\Omega_{k} \right.
\end{eqnarray*}
\begin{eqnarray*}
-\frac{6}{d+3}\Omega_{1}^{1/2}\Omega_{k}-\frac{3d}{d+3}\Omega_{2}
\Omega_{k}
-\frac{3d}{8}\Omega_{2}-\frac{9}{d+3}\Omega_{k}\Omega_{m}+\frac{3d}{4}\Omega_{2}\Omega_{m}
+\frac{A
B(a^{3}b^{d})^{(1+\alpha)(1+\delta)}(d+3)(d+12)(1+\alpha)(1+\delta)^{2}}{8(A(a^{3}b^{d})^{(1+\alpha)(1+\delta)}+B)^{2}}
\end{eqnarray*}
\begin{equation}
\left.+\left(\frac{d}{16}\Omega_{2}+\frac{3}{4}\Omega_{m}-\frac{3}{4(d+3)}\Omega_{k}-\frac{3}{8}\right)
\times \left(
-12+\frac{A(d+12)(1+\delta)}{A+B(a^{3}b^{d})^{-(1+\alpha)(1+\delta)}}
\right)-\left(
-12+\frac{A(d+12)(1+\delta)}{A+B(a^{3}b^{d})^{-(1+\alpha)(1+\delta)}}
\right)\frac{\dot{\Omega}_{x}}{8H}\right]
\end{equation}
These are the expressions for $\{r,s\}$ parameters in terms of
fractional densities for interacting case of generalized Chaplygin
gas model in higher dimensional Cosmology, where
$\dot{\Omega}_{m}$ and $\dot{\Omega}_{x}$ are given
by the equations (31) and (32).\\

\section{Conclusions}

In this work, we have considered $N~(=4+d)$-dimensional Einstein
field equations in which 4-dimensional space-time is described by
a FRW metric and that of the extra $d$-dimensions by an Euclidean
metric. We have calculated the corrections to statefinder
parameters $\{r,s\}$ and deceleration parameter $q$ due to
variable gravitational constant $G$ in higher dimensional
Cosmology. These corrections are relevant because several
astronomical observations provide constraints on the variability
of $G$. We have first assumed that the dark energy do not interact
with dark matter. Next we have considered the dark energy and dark
matter are not separately conserved i.e., they interact with each
other with a particular interacting term in the form $Q=(d+3)
\delta H \rho_{x}$ where $\delta $ is a couple constant. In both
the cases, the statefinder parameters have been found in terms of
the dimensionless density parameters as well as EoS parameter
$\omega$ and the Hubble parameter. An important thing to note is
that these are the $G$-corrected statefinder parameters and they
remain geometrical parameters as previous. Because, the parameter
$\triangle G$ is a pure number and is independent of the geometry.
Finally we have analyzed the above statefinder parameters in terms
of some observable parameters for the non-interacting and
interacting cases when the universe is filled with generalized
Chaplygin gas. These dynamical statefinder parameters may generate
different stages of the anisotropic universe in higher dimensional
Cosmology if the
observable parameters are known for interacting and non-interacting models.\\

\end{document}